\documentclass[12pt]{article}
\usepackage[numbers]{natbib}
\usepackage{a4}
\usepackage{amsmath}
\usepackage{amssymb}
\usepackage{latexsym}
\usepackage{amssymb}
\usepackage{amsthm}
\usepackage{epsfig}
\usepackage{natbib}
\usepackage{bm}
\usepackage{color}
\usepackage{tikz} 
\newtheorem*{theorem}{Theorem}
\def \ii{{\mathrm{i}}}
\def \d{{\mathrm{d}}}

\def \pd{\partial}

\def \e{{\mathrm{e}}}

\begin{document}
\title{{\bf Generalized plane strain embedded in three-dimensional anisotropic elasticity}}
\author{
Markus Lazar~$^\text{a,}$\footnote{
{\it E-mail address:} markus.lazar@kit.edu} 
\ ,
Helmut O.K. Kirchner~$^\text{b,}$\footnote{{\it E-mail address:}
  kirchnerhok@hotmail.com}
\\ \\
${}^\text{a}$ 
Karlsruhe Institute of Technology (KIT),\\
Institute of Engineering Mechanics,\\ 
D-76131 Karlsruhe, Germany\\
$^\text{b}$ INM - Leibniz Institute for New Materials, \\
D-66123 Saarbr\"{u}cken, Germany\\
}

\date{\today}    
\maketitle

\begin{abstract}
The theory of anisotropic generalized plane strain of line forces and dislocation lines is embedded in three-dimensional elasticity of point forces and dislocation densities. Embedding in real space is achieved by slicing in reciprocal space using the projection-slice theorem. 
\\

\end{abstract}

\section{Introduction}

The theory of generalized plane strain, which considers two-dimensional (2D) elastic fields caused by dislocations and line forces, has found its final form 
in the so-called ``6-dimensional integral theory" developed by \citet{BL} and codified by \citet{Bacon} and \citet{Lothe}. 
The formalism connects with previous work by \citet{Lekhn} and \citet{Stroh}, as reviewed comprehensively by \citet{Ting}. 
Its relation to three-dimensional (3D) anisotropic elasticity, where the fields are caused by point forces and dislocation densities \citep{LK13} has never been considered. Both generalized plane strain and three-dimensional elasticity start with the appropriate 2D and 3D differential equations and construct their solutions. 
In 3D, the fields vary like the inverse distance between field and source points, in 2D the dependence is logarithmic plus an angular term. 
Obviously, the 2D fields must be obtainable from the 3D fields by integration along one direction, but 
because of anisotropy evaluation of the necessary integral seems to be impossible in real space.
The objective of this text is to relate the 2D and 3D solutions (not the differential equations) by embedding the former into the latter. 
This is done in reciprocal space using the so-called projection-slice theorem (see, e.g., \citep{Bracewell}).

Usually 3D and 2D problems are treated separately, the differential equation for 2D is obtained by disallowing the partial derivative in one direction,  
 $\pd_i$ with $i = 1, 2, 3$ becomes  $\pd_ i$ with $i = 1, 2$. 
 In electrostatics, for example, the 3D Green function $G^{(3)}(\bm R) = -(4\pi |\bm R|)^{-1}$ 
 and the 2D Green function $G^{(2)}(\bm R) = (2\pi)^{-1} \ln |\bm R|$ 
 are calculated independently of each other. 
 Physically, it is clear that integration of the 3D solutions over one direction must produce the 2D solutions. 
 Such 3D to 2D embedding is more straightforward in Fourier space. 
 Integration over $\d x_3$ and resulting $x_3$-independence amounts to $k_3=0$ and concomitant suppression of the $\d k_3$ integration. 
 The simple example of electrostatics serves to explain the idea. 
 In potential theory, the Green functions for the Laplace operator, 
 \begin{align}
 \label{Lapl}
                   \Delta G^{(3)}(\bm R) = \delta^{(3)}(\bm R)\,,\qquad              
                   \Delta G^{(2)}(\bm R) = \delta^{(2)}(\bm R)\,,
\end{align}                   
are (see \citep{Wl})
\begin{align}
\label{GF-3D}
G^{(3)}(\bm R) &= -\frac{1}{4\pi |\bm R|}\nonumber\\
&=-\frac{1}{(2\pi)^3}
\int_{-\infty}^\infty  \int_{-\infty}^\infty \int_{-\infty}^\infty                                                                              
    \frac{1}{k_1^2+k_2^2 + k_3^2}\,  \e^{\ii(k_1x_1 + k_2x_2 + k_3x_3)} \,\d k_1 \d k_2 \d k_3\,,\\
\label{GF-2D}
G^{(2)}(\bm R) &= \frac{1}{2\pi}\,  \ln |\bm R| \nonumber \\
&=-\frac{1}{(2\pi)^2}
\int_{-\infty}^\infty  \int_{-\infty}^\infty                                                                           
    \frac{1}{k_1^2+k_2^2}\,  \e^{\ii(k_1x_1 + k_2x_2)}\, \d k_1 \d k_2 \,.
\end{align}                                          
In Fourier space, the integrands behave like $k^{-2}$ both in 3D and 2D, 
but the different transforms without and with suppression of $k_3$ produce $-(4\pi |\bm R|)^{-1}$ and $(2\pi)^{-1} \ln |\bm R|$. 
To descend from the 3D solution $G^{(3)}(\bm R)$ in real space, 
one constructs its Fourier transform, uses the projection-slice theorem by putting $k_3 = 0$ in it and transforms back to 
$G^{(2)}(\bm R)$. 
The relation between 3D and 2D differential equations is simple in real space, the relation between 3D and 2D solutions is simple in Fourier space. 
In anisotropic elasticity, the expressions are more complicated, the 2D and 3D solutions look distinctly unrelated, 
their connection is far from obvious, but the same principle prevails. 
The objective of the present paper is the application of the projection-slice theorem to anisotropic elasticity.

\section{Two-dimensional anisotropic elasticity}

\begin{figure}[h]
\centerline{
\begin{tikzpicture} [scale=0.7]
\begin{scope}[line width=1.0pt]
\draw [->] (0,0) -- (5,0,0) node [right] {$\bm m_0$} ; 
\draw [->] (0,0) -- (0,5,0) node [above] {$\bm n_0$}; 
\draw [->] (0,0) -- (4.6,2,0) node [right] {$\bm m$}; 
\draw [->] (0,0) -- (-2,4.6,0) node [above] {$\bm n$}; 
\draw [->] (0,0) -- (4,4.5,0) node [right] {$\bm x$}; 
\draw[->] 
  (2,0) arc (0:23:2cm) node [right] {$\  \phi$} ;
\draw[->] 
  (3,0) arc (0:48:3cm) node [right] {$\quad \omega$} ;
 \end{scope}
\end{tikzpicture}
}
\caption{The unit vectors $\bm m$ and $\bm n$ are to be turned anticlockwise from $\bm m_0$ and $\bm n_0$ by an angle $\phi$, 
and the vector $\bm x$ with angle $\omega$.}
\label{Fig}
\end{figure}
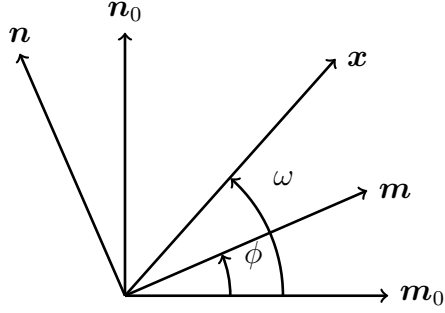

Here, we choose two orthogonal unit vectors $\bm m$ and $\bm n$ which are normal to $\bm t$ such that
$(\bm m,\bm n,\bm t)$ forms a right-handed Cartesian basis. 
This basis is rotated around $\bm t$ by an angle $\phi$ against another fixed $(\bm m_0,\bm n_0,\bm t)$ basis, such that
\begin{align}
\label{mnt}
\begin{pmatrix}
\bm m(\phi)\\
\bm n(\phi)
\end{pmatrix}
&=
 \begin{pmatrix}
\cos\phi & \sin\phi\\
-\sin\phi & \cos\phi
\end{pmatrix}
\begin{pmatrix}
\bm m_0
\\
\bm n_0
\end{pmatrix}
\end{align}
as shown in Fig.~\ref{Fig}.
Moreover, the field vector $\bm x$ reads with respect to the unit basis  $\bm m_0$ and $\bm n_0$
\begin{align}
\label{x-rel}
\bm x&=r(\bm m_0 \cos \omega+\bm n_0 \sin \omega)\,,
\end{align}
where $r=|\bm x|$ (see Fig~\ref{Fig}).

We consider a straight dislocation with Burgers vector $\bm b$ and a line force with strength $\bm F$ located at the origin of the coordinate system.
The line defect runs along the axis $\bm t$. 
The fields of the straight dislocation and line force are the displacement vector $\bm u$ and the stress function vector $\bm \Phi$, 
the sources are the Burgers vector $\bm b$ and the body force  strength $\bm F$. 
Neither $\bm u$  nor $\bm \Phi$ are state quantities. Nevertheless, $\bm u$  and $\bm b$
could be called the geometric field and the geometric source, $\bm \Phi $  and $\bm F$ could be called the physical field and the physical source. 
The geometrical and physical sources contribute to both the geometric and physical fields. 
They are connected by the following equation in integral form \citep{Kirchner87}
\begin{align}
\label{u-phi-1}
\begin{pmatrix}
\bm u(r,\omega)\\
\bm\Phi(r,\omega)
\end{pmatrix}
&=
-\frac{1}{(2\pi)^2}
\left[
 \bm I \ln r
+\int_0^{\omega} \bm N(\phi)\, \d\phi 
\right] \int_0^{2\pi} \bm N(\phi)\, \d\phi\
\begin{pmatrix}
\bm b\\
-\bm F
\end{pmatrix}
\,,
\end{align}
where $\bm I$ is the $6\times 6$ identity matrix.
Moreover, Eq.~\eqref{u-phi-1} can be written in rather compact form as (see \citep{KB76,Ting})
\begin{align}
\label{u-phi-2}
\begin{pmatrix}
\bm u(r,\omega)\\
\bm \Phi(r,\omega)
\end{pmatrix}
&=
-\left[\frac{1}{2\pi}\, 
 \bm I \ln r
+\bm {\widetilde N}(\omega)
\right] 
\bm {\widetilde N}(2\pi)
\begin{pmatrix}
\bm b\\
-\bm F
\end{pmatrix}
\,
\end{align}
with 
\begin{align}
\label{N-1}
\bm {\widetilde N}(\omega)
=\frac{1}{2\pi}
\int_0^{\omega} \bm N(\phi)\, \d\phi \,.
\end{align}
In this manner, Eq.~\eqref{u-phi-2} represents the six-dimensional unification of the displacement fields of a straight dislocation and a line force,
given by \citet{Asaro73}, and  the stress functions of a straight dislocation and a line force, 
given by \citet{Asaro75} in the framework of the integral formalism. 
Eq.~\eqref{u-phi-2} is the field $[\bm u(r,\omega), \bm \Phi(r,\omega)]^T$ 
as function of the sources $[\bm b, -\bm F]^T$ at the origin. 
It was originally obtained in terms of a specifically developed 2D ``integral theory'' without reference to the 3D solution, 
unknown at the time (see \citep{Asaro73,Asaro75,Bacon,Kirchner87,Ting}).

The matrix~\eqref{N-1} fulfills the relation
\begin{align}
\label{N-rel1}
\bm {\widetilde N}(2\pi) \, \bm {\widetilde N}(2\pi)=-\bm I
\,.
\end{align} 
Eq.~\eqref{N-rel1} in conjunction with Eq.~\eqref{u-phi-2} insures the closure condition
\begin{align}
\label{u-phi-rel1}
\begin{pmatrix}
\bm u(r,2\pi)\\
\bm\Phi(r,2\pi)
\end{pmatrix}
-
\begin{pmatrix}
\bm u(r,0\\
\bm\Phi(r,0)
\end{pmatrix}
=
\begin{pmatrix}
\bm b\\
-\bm F
\end{pmatrix}
\,.
\end{align} 
It is interesting to note that Eq.~\eqref{u-phi-1} is the solution of the equation (see \citep{AKT95,Ting})
\begin{align}
\label{u-phi-pde}
\left[\bm I\, \frac{1}{r}\, \pd_\omega-\bm N(\omega)\,\pd_r \right]
\begin{pmatrix}
\bm u(r,\omega)\\
\bm\Phi(r,\omega)
\end{pmatrix}
=\begin{pmatrix}
\bm 0\\
\bm 0
\end{pmatrix}
\,.
\end{align}

The  $6\times 6$ matrix $\bm{N}(\phi)$ is defined by contraction of the elastic constants 
with orthogonal unit vectors $\bm m$ and $\bm n$ according to 
$(m\, m)_{ik} = m_j C_{ijkl} m_l$, 
$(m\, n)_{ik} = m_j C_{ijkl} n_l$, etc. 
As mentioned above,
the vectors $\bm m(\phi)$ and $\bm n(\phi)$ are turned against the $\bm m_0$, $\bm n_0$ coordinate system by an angle $\phi$
(see Fig. \ref{Fig}), 
so that $\bm N$ depends on the angle $\phi$ 
\begin{align}
\label{N}
\bm N(\phi)=
-\begin{pmatrix}
(n\, n)^{-1}\,(n\, m) & (n\,n)^{-1}\\
(m\,n)\,(n\, n)^{-1}\,(n\, m)-(m\, m)\ & (m\, n)\,(n\, n)^{-1}
\end{pmatrix}
\,.
\end{align}
The structure of $\bm N(\phi)$ is such that it varies like a tangent of the real angle $\phi$~\citep{Bacon}, 
though its six complex eigenvalues vary like complex tangents of a complex angle with six imaginary parts, the real part being $\phi$,
\begin{align}
\label{N-tan}
\frac{\d \bm N(\phi)}{\d\phi} + \bm N^2(\phi) + \bm I = \bm 0
\,.
\end{align}

Although the solution~\eqref{u-phi-1} produces the correct closure condition~\eqref{u-phi-rel1}, it is not obvious how it does. 
To clarify the issue, a purely geometric field is extracted by use of Lothe's integral equation (Eq.~\eqref{N-int2} in Appendix)    
\begin{align}
\label{u-phi-3}
\begin{pmatrix}
\bm u(r,\omega)\\
\bm\Phi(r,\omega)
\end{pmatrix}
&=
-\frac{1}{(2\pi)^2}
\left[
 \int_0^{2\pi} \bm N(\phi)\, \d\phi
 \, \ln r
-2\pi\omega\, \bm I
+
 \int_0^{2\pi} \bm N(\phi)\, \ln|\sin(\phi-\omega)|\, \d\phi
 \right]
\begin{pmatrix}
\bm b\\
-\bm F
\end{pmatrix}
\,.
\end{align}
The ``const" in Eq.~\eqref{N-int2} giving a shift of $(\bm u,\bm \Phi)^T$ by a multiple of $(\bm b,-\bm F)^T$ is irrelevant. 
Both in Eqs.~\eqref{u-phi-1} and \eqref{u-phi-3} there are discontinuities at $\omega=2\pi$, 
less obvious in  \eqref{u-phi-1} than in \eqref{u-phi-3}.
There is no discontinuity in the third term  at any value of $\omega$, 
the divergencies of the integrand are integrated over.

\section{Three-dimensional solutions}

For elastic homogeneity of an infinite anisotropic elastic medium
the expressions for  $\bm u(\bm\alpha)$,  $\bm u(\bm f)$ and $\bm \Phi(\bm\alpha)$ are given by \citet{LK13}  
\begin{align}
\label{u-A}
u_i(\bm x)&=\frac{1}{8\pi^2}\int_{\Bbb R^3}\frac{1}{|\bm x-\bm y|}
\int_0^{2\pi}\epsilon_{mnr}C_{lkjn}\,
\kappa_m \kappa_k (\kappa C \kappa)_{il}^{-1}\, \d \phi\, \alpha_{jr}(\bm y)\, \d \bm y
\,,\\
\label{u-f}
u_i(\bm x)&=\frac{1}{8\pi^2}\int_{\Bbb R^3}\frac{1}{|\bm x-\bm y|}
\int_0^{2\pi} (\kappa C \kappa)_{ij}^{-1}\, \d \phi\, f_{j}(\bm y)\, \d \bm y
\,,\\
\label{phi-A}
\Phi_{it}(\bm x)&=\frac{1}{8\pi^2}\int_{\Bbb R^3}\frac{1}{|\bm x-\bm y|}
\int_0^{2\pi}
\epsilon_{tkl}C_{ikmn}\epsilon_{npr}C_{qsjp} 
\kappa_l \kappa_s (\kappa C \kappa)_{mq}^{-1}\, \d \phi \,
\alpha_{jr}(\bm y) \, \d \bm y
\nonumber\\
&=
\frac{1}{4\pi}\int_{\Bbb R^3}\frac{1}{|\bm x-\bm y|}\,
\big(
C_{iljl} \delta_{tr} -C_{irjt}\big)\,
\alpha_{jr}(\bm y) \, \d \bm y
\nonumber\\
&\quad
-\frac{1}{8\pi^2}\int_{\Bbb R^3}\frac{1}{|\bm x-\bm y|}
\int_0^{2\pi}
\big(
C_{ikml}C_{jkqs}\delta_{tr} -C_{irml}C_{jtqs}\big)
\kappa_l \kappa_s (\kappa C \kappa)_{mq}^{-1}\, \d \phi \,
\alpha_{jr}(\bm y) \, \d \bm y
\,,
\end{align}
where $\bm \alpha$ and $\bm f$ denote the dislocation density tensor and the body force vector, respectively.
$\bm y$ is the position of the sources. 
Reciprocity is respected in \eqref{u-f} and \eqref{phi-A}, $u_i(\bm x) f_i(\bm y)=u_i(\bm y) f_i(\bm x)$ 
and  $\alpha_{ij}(\bm x) \Phi_{ij}(\bm y)=\alpha_{ij}(\bm y) \Phi_{ij}(\bm x)$. 
In reality, the one-dimensional integrations in Eqs.~\eqref{u-A}--\eqref{phi-A} are three-dimensional Fourier transforms, 
reduced by the theorem \citep{LR}  valid for functions homogeneous in $k^{-2}$ and $|\bm x-\bm y|^{-1}$, 
which is here the case.
The unit vectors $\bm \kappa_m(\phi)$, normal to $(\bm x-\bm y)$, are turned around the direction of $(\bm x-\bm y)$. 

The expression \eqref{u-A} is not the complete displacement field caused by a dislocation loop, 
a purely geometric contribution, $u_i = -b_i \Omega/4\pi$, must be added (see \citep{LK13}).
 $\Omega$ is the solid angle under which the dislocation loop is seen. 
The set of equations \eqref{u-A}--\eqref{phi-A} lacks the stress function tensor caused by a body force. 
It is constructed from the fact \citep{LK13} that the Laplacian of the stress function $\bm \Phi$ is the curl of the stress $\bm\sigma$,
$\Delta \bm \Phi=\text{curl}\, \bm \sigma=\text{curl}(\bm C:\text{grad}\, \bm u)$.
With $\bm u(\bm f)$ in Eq. \eqref{u-f} the result reads as 
\begin{align}
\label{phi-f}
\Phi_{it}(\bm x)&=\frac{1}{8\pi^2}\int_{\Bbb R^3}\frac{1}{|\bm x-\bm y|}\int_0^{2\pi}
\epsilon_{tkl}C_{ikmn}
\kappa_l \kappa_n (\kappa C \kappa)_{mj}^{-1}\, \d \phi\,  f_{j}({\bm y})\, \d \bm y
\,.
\end{align}

\section{Projection-slice theorem: 3D $\rightarrow$ 2D projection}

The projection-slice theorem (alternatively, the central-slice theorem or the Fourier-slice theorem) is a powerful tool 
for the analysis of medical computed tomography (CT) scans where a ``projection'' is an X-ray image of an internal organ. 
It was originally derived in the context of radio astronomy in 1956 (see, e.g., \citep{Bracewell}). 
The  projection-slice theorem has also been used in acoustic tomography~\citep{Santosa} and imaging anisotropic media~\citep{Chinta}.
In this work, we will use, for the first time, the projection-slice theorem in 3D anisotropic dislocation theory.

\begin{theorem}
In 3 dimensions, 
the {\bf projection-slice theorem} states that the  2-dimensional Fourier transform of the {\bf projection} of a 3-dimensional  function $f(x_1,x_2,x_3)$ 
onto a 2-dimensional linear submanifold is equal to a 2-dimensional {\bf slice} of the 3-dimensional Fourier transform of that function 
consisting of a 2-dimensional linear submanifold through the origin in the Fourier space which is parallel to the projection submanifold.
In operator terms, if
\begin{itemize}
\item
    ${\cal F}_{2}$ and ${\cal F}_{3}$ are the 2- and 3-dimensional Fourier transform operators,
\item
    ${\cal P}_{2}$ is the projection operator (which projects a 3D function onto a 2D plane),
\item
    ${\cal S}_{2}$ is a slice operator (which extracts a 2D central slice from a 3D function),
\end{itemize}
then it holds
\begin{align}
\label{PST}
  {\cal F}_{2}\,{\cal P}_{2}={\cal S}_{2} \,{\cal F}_{3}\,. 
\end{align}
\end{theorem}

Using the 2-dimensional inverse Fourier transform ${\cal F}^{-1}_{2}$ with ${\cal F}^{-1}_{2}\, {\cal F}_{2}=1$
on the projection-slice theorem~\eqref{PST}, 
the 2-dimensional projection can be written 
\begin{align}
\label{P}
  {\cal P}_{2}={\cal F}^{-1}_{2}\,{\cal S}_{2} \,{\cal F}_{3}\,,
\end{align}
which consists of the three steps: ${\cal F}_{3}$, ${\cal S}_{2}$ and ${\cal F}^{-1}_{2}$.

The projection of the 3D function  $f(x_1,x_2,x_3)$ onto the $x_1 x_2$ plane 
is given by the 2D function  $f(x_1,x_2)$:
\begin{align}
\label{f-P}
f(x_1,x_2)&={\cal P}_{2}\, f(x_1,x_2,x_3)\nonumber\\
	&=\int_{-\infty}^\infty  f(x_1,x_2,x_3)\, \d x_3\,,
\end{align}	
which is a 1D integral in the real space. 
The 3D Fourier transform of the 3D function  $f(x_1,x_2,x_3)$ is given by 
\begin{align}
\label{f-FT-2}
\widehat{f}(k_1,k_2,k_3)&={\cal F}_{3}\, f(x_1,x_2,x_3)\nonumber\\
	&=\int_{-\infty}^\infty  \int_{-\infty}^\infty \int_{-\infty}^\infty 
	f(x_1,x_2,x_3)\, \e^{-\ii(k_1 x_1 +k_2 x_2 +k_3 x_3)}\,
	\d x_1 \d x_2 \d x_3\,.
\end{align}
The 2D Fourier transform of the 2D function  $f(x_1,x_2)$ reads as
\begin{align}
\label{f-FT-3}
\widehat{f}(k_1,k_2)&={\cal F}_{2}\, f(x_1,x_2)\nonumber\\
	&=\int_{-\infty}^\infty  \int_{-\infty}^\infty \
	f(x_1,x_2)\, \e^{-\ii(k_1 x_1 +k_2 x_2 )}\,
	\d x_1 \d x_2\,.
\end{align}

The 2D slice of $\widehat f(k_1,k_2,k_3)$ at $k_3=0$ is given by
\begin{align}
\label{f-S-2}
\widehat{f}(k_1,k_2,k_3=0)&={\cal S}_{2}\, \widehat f(k_1,k_2,k_3)\nonumber\\
	&=\int_{-\infty}^\infty  \int_{-\infty}^\infty \int_{-\infty}^\infty 
	f(x_1,x_2,x_3)\, \e^{-\ii(k_1 x_1 +k_2 x_2 )}\,
	\d x_1 \d x_2 \d x_3
	\nonumber\\
	&=\int_{-\infty}^\infty  \int_{-\infty}^\infty 
	\left[\int_{-\infty}^\infty 
	f(x_1,x_2,x_3)\, \d x_3\right] \e^{-\ii(k_1 x_1 +k_2 x_2 )}\,
	\d x_1 \d x_2 
	\nonumber\\
	&=\int_{-\infty}^\infty  \int_{-\infty}^\infty 
	 f(x_1,x_2)\, \e^{-\ii(k_1 x_1 +k_2 x_2 )}\,
	\d x_1 \d x_2 
	\nonumber\\
	&={\cal F}_{2}\, f(x_1,x_2)
	\,,
\end{align}
which is nothing but the 2D Fourier transform of the function $f(x_1,x_2)$.

The inverse 2D Fourier transform of the 2D function  $\widehat f(k_1,k_2)$ is
defined by 
\begin{align}
\label{f-inv-FT-2}
f(x_1,x_2)&={\cal F}^{-1}_{2}\, \widehat f(k_1,k_2,)\nonumber\\
	&=\frac{1}{(2\pi)^2}\int_{-\infty}^\infty  \int_{-\infty}^\infty \
	\widehat f(k_1,k_2)\, \e^{\ii(k_1 x_1 +k_2 x_2 )}\,
	\d k_1 \d k_2 \,.
\end{align}

Using Eq.~\eqref{P} with Eqs.~\eqref{f-FT-3}, \eqref{f-S-2} and \eqref{f-inv-FT-2},
the projection~\eqref{f-P} can be written as   
\begin{align}
\label{f-P-2}
f(x_1,x_2)
	&={\cal F}^{-1}_{2}\,{\cal S}_{2} \,{\cal F}_{3}\, f(x_1,x_2,x_3)\nonumber\\
	&={\cal F}^{-1}_{2}\, \widehat f(k_1,k_2,k_3=0)\nonumber\\
	&=\frac{1}{(2\pi)^2}\int_{-\infty}^\infty  \int_{-\infty}^\infty \
	\widehat f(k_1,k_2,k_3=0)\, \e^{\ii(k_1 x_1 +k_2 x_2 )}\,
	\d k_1 \d k_2 \,,
\end{align}
which is a 2D integral in the Fourier space.
Therefore, 
using Eq.~\eqref{f-P-2},
the 2D projection of the 3D function  $f(x_1,x_2,x_3)$ in the $x_3$-direction can be simply computed by 
means of the 2D inverse Fourier transform of the slice  $\widehat f(k_1,k_2,k_3=0)$ 
by substituting $k_3=0$ in the 3D Fourier transform $\widehat f(k_1,k_2,k_3)$.

\section{Projection-slice theorem in 3D anisotropic elasticity}

In 3D isotropic elasticity, the projection~\eqref{f-P} can be easily computed in real space. 
In fact, it was used in the calculation of the displacement fields of straight screw and edge dislocations  
from the 3D fields in isotropic elasticity by \citet{Mura}.
However, in 3D anisotropic elasticity, the projection using Eq.~\eqref{f-P} is difficult and inconvenient.
Therefore, in 3D anisotropic elasticity, the projection using Eq.~\eqref{f-P-2} is the proper tool
for the projection of 3D anisotropic elasticity toward 2D and will be used in this section.

The 2D fields can be obtained by using the projection given in Eq.~\eqref{f-P}
\begin{align}
\label{u-P}
u_i(x_1,x_2)&={\cal P}_2\big[u_i(x_1,x_2,x_3)\big]\,,\\
\label{Phi-P}
\Phi_{it}(x_1,x_2)&={\cal P}_2\big[\Phi_{it}(x_1,x_2,x_3)\big]
\end{align}
and by using the projection-slice theorem via Eq.~\eqref{f-P-2}
\begin{align}
\label{u-P-2}
u_i(x_1,x_2)&={\cal F}^{-1}_2 {\cal S}_2 \,{\cal F}_3\big[u_i(x_1,x_2,x_3)\big]\nonumber\\
	            &={\cal F}^{-1}_2 \big[\widehat u_i(k_1,k_2,k_3=0)\big]\,,\\
\label{Phi-P-2}
\Phi_{it}(x_1,x_2)&={\cal F}^{-1}_2 {\cal S}_2 \,{\cal F}_3\big[\Phi_{it}(x_1,x_2,x_3)\big]\nonumber\\
	            &={\cal F}^{-1}_2 \big[\widehat \Phi_{it}(k_1,k_2,k_3=0)\big]\,.
\end{align}
In the following, we will use Eqs. \eqref{u-P-2} and \eqref{Phi-P-2}
for the projection  of the 3D fields~\eqref{u-A}--\eqref{phi-f}  toward 2D fields.

\subsection{First step: 3D Fourier transform}
The 3D Fourier transforms of Eqs.~\eqref{u-A}--\eqref{phi-f} read as
\begin{align}
\label{u-A-FT}
\widehat u_i(\bm k)&=-\frac{1}{k^2}\,
\epsilon_{mnr}C_{lkjn}\,
\kappa_m \kappa_k (\kappa C \kappa)_{il}^{-1}\, 
\widehat\alpha_{jr}(\bm k)
\,,\\
\label{u-f-FT}
\widehat u_i(\bm k)&= \frac{1}{k^2}\, (\kappa C \kappa)_{ij}^{-1}\, \widehat f_{j}(\bm k)
\,,\\
\label{phi-A-FT}
\widehat \Phi_{it}(\bm k)&=-\frac{1}{k^2}\,
\epsilon_{tkl}C_{ikmn}\epsilon_{npr}C_{qsjp} 
\kappa_l \kappa_s (\kappa C \kappa)_{mq}^{-1}\, \widehat \alpha_{jr}(\bm k)
\,,\\
\label{phi-f-FT}
\widehat \Phi_{it}(\bm k)&=-\frac{1}{k^2}\,
\epsilon_{tkl}C_{ikmn}
\kappa_l \kappa_n (\kappa C \kappa)_{mj}^{-1}\, \widehat  f_{j}({\bm k})
\,,
\end{align}
where $\bm k\in\Bbb R^3$ and $k=\sqrt{k_1^2+k_2^2+k_3^2}$.

\subsection{Second step: 2D slice at $k_3=0$}

In Fourier space, the slices of Eqs. \eqref{u-A-FT}--\eqref{phi-f-FT} at $k_3=0$ become
\begin{align}
\label{u-A-S}
\widehat u_i(k_1,k_2)&=-\frac{1}{k^2}\,
\epsilon_{mnr}C_{lkjn}\,
\kappa_m \kappa_k (\kappa C \kappa)_{il}^{-1}\, 
\widehat\alpha_{jr}(k_1,k_2)
\,,\\
\label{u-f-S}
\widehat u_i(k_1,k_2)&= \frac{1}{k^2}\, (\kappa C \kappa)_{ij}^{-1}\, \widehat f_{j}(k_1,k_2)
\,,\\
\label{phi-A-S}
\widehat \Phi_{it}(k_1,k_2)&=-\frac{1}{k^2}\,
\epsilon_{tkl}C_{ikmn}\epsilon_{npr}C_{qsjp} 
\kappa_l \kappa_s (\kappa C \kappa)_{mq}^{-1}\, \widehat \alpha_{jr}(k_1,k_2) 
\,,\\
\label{phi-f-S}
\widehat \Phi_{it}(k_1,k_2)&=-\frac{1}{k^2}\,
\epsilon_{tkl}C_{ikmn}
\kappa_l \kappa_n (\kappa C \kappa)_{mj}^{-1}\, \widehat  f_{j}(k_1,k_2)
\,,
\end{align}
where $\bm k\in\Bbb R^2$, $k=\sqrt{k_1^2+k_2^2}$ and $\bm \kappa=\bm k/k\in\Bbb R^2$. 

Using the fact that $\bm \kappa\in\Bbb R^2$ and the major symmetry of the elastic tensor $C_{ijkl}=C_{klij}$, 
Eqs.~\eqref{u-A-S}, \eqref{phi-A-S} and \eqref{phi-f-S} can be written
\begin{align}
\label{u-A-S-2}
\widehat u_i(k_1,k_2)&=-\frac{1}{k^2}\,
\epsilon_{3\alpha\beta}C_{j\alpha l\gamma}\,
\kappa_\beta \kappa_\gamma (\kappa C \kappa)_{il}^{-1}\, 
\widehat\alpha_{j3}(k_1,k_2)
\,,\\
\label{phi-A-S-2}
\widehat \Phi_{i3}(k_1,k_2)&=-\frac{1}{k^2}\,
\epsilon_{3\beta\alpha}C_{i\alpha l\gamma }\epsilon_{3\gamma\nu}C_{r\mu j\nu} 
\kappa_\beta \kappa_\mu (\kappa C \kappa)_{lr}^{-1}\, \widehat \alpha_{j3}(k_1,k_2) 
\,,\\
\label{phi-f-S-2}
\widehat \Phi_{i3}(k_1,k_2)&=-\frac{1}{k^2}\,
\epsilon_{3\beta\alpha}C_{i\alpha l\gamma}
\kappa_\beta \kappa_\gamma (\kappa C \kappa)_{lj}^{-1}\, \widehat  f_{j}(k_1,k_2)
\,,
\end{align}
where the Latin indices take the values of 1, 2 and 3, whereas the Greek indices take the values of 1 and 2 only. 
In 2D, the indices $t$ and $r$ became 3.  
There remain 3 displacement components $u_i=u_i(x_1,x_2)$ and 3 stress functions $\Phi_{i3}=\Phi_{i3}(x_1,x_2)$. 

Now, Eqs. \eqref{u-A-S-2},   \eqref{u-f-S},  \eqref{phi-A-S-2} and \eqref{phi-f-S-2}
can be written in the following matrix equation form
\begin{align}
\label{u-phi-M-1}
\begin{pmatrix}
\widehat u_{i}(k_1,k_2)\\
\widehat \Phi_{i3}(k_1,k_2)
\end{pmatrix}
&=
-\frac{1}{k^2} 
 \begin{pmatrix}
 \epsilon_{3\alpha\beta} \kappa_\beta C_{j\alpha l\gamma} \kappa_\gamma  (\kappa\,\kappa)^{-1}_{il} 
& (\kappa\, \kappa)^{-1}_{ij}\\
\epsilon_{3\beta\alpha}C_{i\alpha l\gamma}\epsilon_{3\gamma\nu}C_{r\mu j\nu} \kappa_\beta \kappa_\mu (\kappa\, \kappa)^{-1}_{lr}
& \epsilon_{3\alpha\beta} \kappa_\beta C_{i\alpha l\gamma} \kappa_\gamma(\kappa\,\kappa)^{-1}_{lj} 
\end{pmatrix}
\begin{pmatrix}
\widehat\alpha_{j3}(k_1,k_2)
\\
-\widehat f_{j}(k_1,k_2)
\end{pmatrix}
\,.
\end{align}

Identifying $\bm \kappa=\bm n$ and 
$\epsilon_{3\alpha\beta} \kappa_\beta=\epsilon_{3\alpha\beta} n_\beta=m_\alpha$ (see Fig. \ref{Fig}), 
the north-west (NW) block of the matrix in  Eq.~\eqref{u-phi-M-1} becomes
\begin{align}
\label{NW}
\epsilon_{3\alpha\beta} \kappa_\beta C_{j\alpha l\gamma} \kappa_\gamma  (\kappa\,\kappa)^{-1}_{il} 
=m_\alpha C_{j\alpha l\gamma} n_\gamma  (n\,n)^{-1}_{il} 
=(m\,n)_{jl}(n\,n)^{-1}_{il}=(n\,n)^{-1}_{il}(n\,m)_{lj}\,,
\end{align}
which is the NW block of $\bm N$ given in Eq. \eqref{N}. 
The north-east (NE) block of the matrix in Eq.~\eqref{u-phi-M-1} is simply $(\kappa\, \kappa)^{-1}_{ij}=(n\, n)^{-1}_{ij}$, 
being the NE block in Eq. \eqref{N}. 
The south-east (SE) block of the matrix in Eq.~\eqref{u-phi-M-1} becomes
\begin{align}
\label{SE}
\epsilon_{3\alpha\beta} \kappa_\beta C_{i\alpha l\gamma} \kappa_\gamma(\kappa\,\kappa)^{-1}_{lj} 
=m_\alpha C_{i\alpha l\gamma} n_\gamma(n\,n)^{-1}_{lj} 
=(m\, n)_{il} (n\,n)^{-1}_{lj} \,,
\end{align}
which is the SE block of $\bm N$ given in Eq. \eqref{N}. 
The south-west (SW) block  of the matrix in Eq.~\eqref{u-phi-M-1} becomes
\begin{align}
\label{SW}
&\epsilon_{3\beta\alpha}C_{i\alpha l\gamma}\epsilon_{3\gamma\delta}C_{r\mu j\delta}\,\kappa_\beta \kappa_\mu (\kappa\, \kappa)^{-1}_{lr}
=C_{i\alpha l\beta}n_\beta C_{j\alpha r\mu} n_\mu (n\, n)^{-1}_{lr}-C_{i\alpha j\alpha}\nonumber\\
&\hspace{3cm}
=(m\,n)_{il}(m\, n)_{jr}(n\,n)^{-1}_{lr}
+(n\,n)_{il}(n\, n)_{jr}(n\,n)^{-1}_{lr}
-(m\,m)_{ij}-(n\,n)_{ij}\nonumber\\
&\hspace{3cm}
=(m\,n)_{il}(n\,n)^{-1}_{lr}(n\, m)_{rj}-(m\,m)_{ij}\,,
\end{align}
being the SW block of $\bm N$ given in Eq. \eqref{N}. 
In this way, the matrix equation \eqref{u-phi-M-1} can be written in terms of 
the $6\times 6$ matrix $\bm N(\phi)$ given in Eq.~\eqref{N} 
\begin{align}
\label{u-phi-M-2}
\begin{pmatrix}
\widehat{\bm u}(k_1,k_2)\\
\widehat{\bm \Phi}(k_1,k_2)
\end{pmatrix}
&=
\frac{1}{k^2} \, \bm N(\phi)
\begin{pmatrix}
\widehat{\bm \alpha}(k_1,k_2)
\\
-\widehat{\bm f}(k_1,k_2)
\end{pmatrix}
\,.
\end{align}
In the Fourier space, the dislocation density and the line force density 
of straight line defects are given  by
$\widehat{\bm \alpha}=\bm b$ and 
$\widehat{\bm f}=\bm F$,
which are the Fourier transforms of $\bm \alpha(x_1,x_2)=\bm b\, \delta(x_1)\delta(x_2)$
and  $\bm f(x_1,x_2)=\bm F\, \delta(x_1)\delta(x_2)$,
and the matrix equation \eqref{u-phi-M-2} becomes for straight line defects 
\begin{align}
\label{u-phi-M-3}
\begin{pmatrix}
\widehat{\bm u}(k_1,k_2)\\
\widehat{\bm \Phi}(k_1,k_2)
\end{pmatrix}
&=
\frac{1}{k^2} \, \bm N(\phi)
\begin{pmatrix}
\widehat{\bm b}
\\
-\widehat{\bm F}
\end{pmatrix}
\,.
\end{align}
This equation clarifies the meaning of the $6\times 6$ matrix $\bm N(\phi)$, 
which is the central quantity of the integral theory: 
In reciprocal space it is the coefficient between fields and sources, 
only the 2D inverse Fourier transform produces the characteristic logarithmic plus angular dependence.

\subsection{Third step: 2D inverse Fourier transform}

The 2D inverse Fourier transform of Eq. \eqref{u-phi-M-3} reads in Cartesian coordinates 
\begin{align}
\label{u-phi-M-4}
\begin{pmatrix}
\bm u(x_1,x_2)\\
\bm \Phi_3(x_1,x_2)
\end{pmatrix}
&=
\frac{1}{(2\pi)^2}\int_{-\infty}^\infty  \int_{-\infty}^\infty \,
\frac{1}{k^2} \, \bm N(\phi)\, 
 \e^{\ii(k_1 x_1 +k_2 x_2 )}\,
	\d k_1 \d k_2 \,
\begin{pmatrix}
\bm b
\\
-\bm F
\end{pmatrix}
\,
\end{align}
and in polar coordinates it becomes
\begin{align}
\label{u-phi-M-5}
\begin{pmatrix}
\bm u(x_1,x_2)\\
\bm \Phi(x_1,x_2)
\end{pmatrix}
&=
\frac{1}{(2\pi)^2}\int_{0}^{2\pi}
\bm N(\phi)
 \int_{0}^\infty \,
\frac{1}{k} \, \e^{\ii k\,  \bm n \cdot \bm x}\,
	\d k\, \d \phi \,
\begin{pmatrix}
\bm b
\\
-\bm F
\end{pmatrix}
\,.
\end{align}
The $k$-integration can be carried out by using the 
principal value integral (see Eq. (32) in Sec. 9 in \citep{Wl}, and Eq. (33c) in Sec. 6.4 in \citep{Kanwal})
\begin{align}
\label{k-int}
\frac{1}{2}\, {\cal {P}}
\int_{-\infty}^\infty 
 \frac{1}{k}\, \e^{\ii k t}\, \text{d}k 
= {\cal {P}}
\int_{0}^\infty 
 \frac{1}{k}\, \e^{\ii k t}\, \text{d}k 
 =-\gamma-\ln |t|\,,
\end{align}
where $\gamma$ denotes the Euler constant
and ${\cal {P}}$ means the principal value. 
Another expression for the integral in Eq. \eqref{k-int} is the limit $z\rightarrow  0$  of the exponential integral $E_1(z)$, 
formula 5.1.1 of \citet{Abram}.
The constant term $\gamma$ can be neglected because it gives just an irrelevant constant displacement and stress function.
Using Eq.~\eqref{k-int}, Eq. \eqref{u-phi-M-5} reduces to 
\begin{align}
\label{u-phi-M-6}
\begin{pmatrix}
\bm u(x_1,x_2)\\
\bm \Phi(x_1,x_2)
\end{pmatrix}
&=
-\frac{1}{(2\pi)^2}\int_{0}^{2\pi}
\bm N(\phi)
 \,\ln |\bm x \cdot \bm n |\,
	\d \phi \,
\begin{pmatrix}
\bm b
\\
-\bm F
\end{pmatrix}
\,.
\end{align}

For the geometry given in Fig. \ref{Fig} and using Eqs. \eqref{mnt} and \eqref{x-rel},
the inner product between the vectors $\bm x$ and $\bm n$ reads as
\begin{align}
\bm x\cdot\bm n=r\sin(\omega-\phi)=-r \sin(\phi-\omega)\,,
\end{align}
so that the $\ln$-term in Eq.~\eqref{u-phi-M-6} can be written as 
\begin{align}
\label{ln}
\ln|\bm x \cdot\bm n|
=\ln r +\ln |\sin(\phi-\omega)|\,. 
\end{align}
Substituting Eq. \eqref{ln} into Eq. \eqref{u-phi-M-6}, we obtain in polar coordinates
\begin{align}
\label{u-phi-M-7}
\begin{pmatrix}
\bm u(r,\omega)\\
\bm \Phi(r,\omega)
\end{pmatrix}
&=-\frac{1}{(2\pi)^2}
\left[
 \int_0^{2\pi} \bm N(\phi)\, \d\phi
 \, \ln r
+
 \int_0^{2\pi} \bm N(\phi)\, \ln|\sin(\phi-\omega)|\, \d\phi
 \right]
\begin{pmatrix}
\bm b\\
-\bm F
\end{pmatrix}
\,.
\end{align}
This is the non-$\omega$ part of the generalized plane strain solution \eqref{u-phi-3}, it completes the embedding.
The geometric $\omega$ part in Eq.~\eqref{u-phi-3} is due to the defect-topology of the line defect and its 3D version.
For instance, the purely geometric term $\Omega$ of a dislocation loop in 3D becomes $-2\omega$ of a straight dislocation in 2D 
(see Appendix, Eq. \eqref{omega2}).

\section{Conclusion}

Hitherto the field equations of generalized plane strain and the 3D field equations were solved separately, the relation between their solution remained uninvestigated. 
The question, how the fields of generalized plane strain are imbedded into the 3D fields, remained open. 
The answer, if such a reduction of the 3D to the 2D solution is possible, is given in the affirmative. 
Generalized plane strain is the projection of the 3D solution into 2D.  
This projection, impossible in real space, is achieved by slicing in reciprocal space by means of the projection-slice theorem~\citep{Bracewell}.
 A relative of the Fourier-slice theorem, the Radon transform, was used by~\citet{Bacon} for calculating derivatives of the Green function. 
 However, the Radon transform, $\cal R$, is the integral transform which takes a function $f$ defined on a plane to a function ${\cal{R}}[f]$ defined on the 
 2D space of lines in the plane, whose value at a particular line is equal to the line integral over the function over that line. 
 The inversion of ${\cal{R}}[f]$ to $f$ is possible.

Extension of the presented sextic ($6\times 6$) formalism of anisotropic elasticity to the decadic ($10 \times 10$) formalism 
of piezoelectric piezomagnetic elastic media \citep{AKT95} 
and piezoelectric-piezomagnetic-magnetoelectric elastic media \citep{KA96} is possible and straightforward. 

The projection takes the 3D situation with point forces and dislocation densities to the corresponding 2D one, in both the medium is of infinite extent, no boundary conditions are involved. Other 2D situations, like plane strain or plane stress have no 3D equivalent. Because they are already constrained by geometric conditions, like invariance of components or vanishing derivatives in one direction, they are by definition specifically 2D geometries. Finite 2D problems with boundary conditions have no nontrivial 3D equivalent either. 
  
In medical imaging, the projection of the 3D situation is a 2D X-ray picture taken. 
In medicine, the inverse tomographic reconstruction 
of the 3D situation from 2D projections taken in many different direction is of concern. 
The elastic analogue of reconstructing 3D solutions from 2D integral solutions on many different planes remains an interesting, albeit artificial problem. 

The slicing of the Fourier-slice theorem
 elucidates several features of the 2D vs 3D situation. 
For example, the vanishing trace of the 3D stress function, trace$(\bm \Phi) = 0$ in 3D, corresponds to zero divergence, $\text{div}\, \bm \Phi = 0$,
 of the stress vector in 2D. 
The embedding clarifies the nature of the so-called six-dimensional integral theory, it is a 2D inverse Fourier transform obtained as slice of a 3D Fourier transform. 
The ``integral formalism" is more natural than the preceding Stroh and Lekhnitzki theories formulated in terms of eigenvectors and eigenvalues,
 these are spectral representations of the ``integral theory" in Fourier space.

\section*{Acknowledgement}
Markus Lazar gratefully acknowledges the grant from the 
Deutsche Forschungsgemeinschaft (Grant number LA1974/4-2).

\begin{appendix}

\label{appendixA}
\setcounter{equation}{0}
\renewcommand{\theequation}{\thesection.\arabic{equation}}

\section{Lothe's integral equation }

The eigenvalues of $\bm N(\omega)$ behave like tangents of a complex angle, 
$p_\alpha(\omega) = \tan(\Psi_\alpha - \omega)$ (see Eq.~3.5.34 in \citep{Bacon}) 
with negative sign for the real angle. 
Apply the sum formula for the tangent of the complex argument ($\Psi_\alpha - \omega$) 
to two of these.  
The integrals are taken in the principal value sense
\begin{align}
\label{L1}
[1 + p_\alpha(\phi)\,  p_\alpha(\omega)] = -[p_\alpha(\phi) - p_\alpha(\omega)] \cot(\phi - \omega)\,,                                    
\end{align}
where $\alpha=1,\dots,6$.
Integrate over $\d\phi$ from 0 to $2\pi$
\begin{align}
\label{L2}
2\pi+\int_0^{2\pi} p_\alpha(\phi) \, \d\phi\,  p_\alpha(\omega)  
&= -\int_0^{2\pi} p_\alpha(\phi) \cot(\phi-\omega)\, \d \phi
 + \int_0^{2\pi} p_\alpha(\omega)  \cot(\phi-\omega)\,\d\phi    \nonumber\\
   &=-\int_0^{2\pi} p_\alpha(\phi) \cot(\phi-\omega)\, \d \phi\,.                         
\end{align}
The arguments of $p_\alpha(\omega)$ and $p_\alpha(\phi)$ have the same complex part, 
therefore the cotangent in \eqref{L2} has a purely real argument. The second integral is zero. 
One more integration over $\d\omega$ from 0 to $2\pi$ gives a central result of the integral theory
\begin{align}
\label{L3}
\left[\int_0^{2\pi} p_\alpha(\phi) \, \d\phi\right]^2=(2\pi)^2
\,.
\end{align}
Multiplication with the $\omega$-independent eigenvectors of $\bm N$ confirms Eq. \eqref{N-rel1} above,
\begin{align}
\label{L4}
\left[\int_0^{2\pi} \bm N(\phi) \, \d\phi\right]^2=(2\pi)^2 \bm I
\,.
\end{align}
Multiplied with the $\omega$-independent eigenvectors of $\bm N$ to obtain the spectral form, 
the integral equation~\eqref{L2} for the tangent of a complex argument becomes an integral equation for a matrix that varies 
like a tangent (see  Eq.~(114) in \citet{Lothe})
\begin{align}
\label{N-int}
-\frac{1}{2\pi}\, {\cal{P}}\int_0^{2\pi} \bm N(\phi)
\cot(\phi-\omega)\, \d\phi=
\bm  I 
+\bm N(\omega) \,    \int_0^{2\pi}  \bm N(\phi)\, \d\phi
\,.
\end{align}
Integrated over $\d\omega$ 
\begin{align}
\label{N-int2}
\int_0^{2\pi} \bm N(\phi)
\ln|\sin(\phi-\omega)|\, \d\phi=2\pi  \omega\bm  I 
+\left(\int_0^\omega  \bm N(\phi) \, \d\phi\right) \int_0^{2\pi}  \bm N(\phi)\, \d\phi
+\text{const}
\,.
\end{align}        
For no value of $\omega$ a discontinuity is present in Eq.~\eqref{N-int2}. 
The discontinuities at $\omega = 2\pi$ of both the first and second term on the right-hand side cancel, 
the term on the left-hand side is just an integral for any value of $\omega$ 
has no discontinuity for any $\omega$.

\section{Projection of the 3D solid angle $\Omega$}
\setcounter{equation}{0}

The 3D solid angle reads for a dislocation loop (see \citep{LK13,Leibfried,Lothe})
\begin{align}
\label{Omega}
\Omega(x_1,x_2,x_3)=-\int_S\pd_m \frac{1}{R}\, \d S'_m\,,
\end{align}
subtended by the loop area $S'_m$.

For a straight dislocation with dislocation line direction along the $x_3$-axis, $S$ can be chosen as a semi-infinite plane $x'_1x'_3$ for $x'_1>0$ at $x'_2=0$ with  $\d S'_2=\d x'_1\,\d x'_3$ and $m=2$,
the solid angle becomes 
\begin{align}
\label{omega1}
\Omega(x_1,x_2)
=  \int_0^\infty \int_{-\infty}^{\infty}\frac{x_2}{R^3}\, \d x'_3\,  \d x'_1 
\,,
\end{align}
where $R=[(x_1-x'_1)^2+x_2^2+(x'_3)^2]^{1/2}$. 
After integration over $\d x'_3$
\begin{align}
 \int_{-\infty}^{\infty}\frac{1}{R^3}\, \d x'_3 =\frac{2}{(x_1-x'_1)^2+x_2^2}\,,
 \end{align}
 and further integration over $\d x'_1$ the solid angle \eqref{omega1} reduces to (see also \citep{Leibfried,Balluffi})
 \begin{align}
 \label{omega2}
\Omega(x_1,x_2)&=  2\int_0^\infty \frac{x_2}{(x_1-x'_1)^2+x_2^2} \, \d x'_1 
\nonumber\\
&=-2\arctan \left(\frac{x_2}{x_1}\right)
\nonumber\\
&=-2\, \omega(x_1,x_2)\,,
\end{align}
which is nothing but twice the negative polar angle $\omega$.
The $4\pi$ discontinuity of $\Omega$ becomes a $2\pi$ discontinuity of $\omega$. 
The purely geometric part of the displacement field of a straight dislocation is
 \begin{align}
\label{u-geo}
\bm u = -\frac{\bm b\,\Omega}{4\pi}
=\frac{\bm b\,\omega}{2\pi}
\,
\end{align} 
as in Eq.~\eqref{u-phi-3}.
   
\end{appendix}

\end{document}